\documentclass[times,twocolumn,final]{elsarticle}

\usepackage{medima}
\usepackage{framed,multirow}

\usepackage{amssymb}
\usepackage{latexsym}

\usepackage{xurl}

\usepackage{xcolor}

\usepackage{hyperref}

\definecolor{newcolor}{rgb}{.8,.349,.1}

\usepackage[utf8]{inputenc}
\usepackage{subfig}
\usepackage{amsmath}
\usepackage{booktabs}       
\usepackage{makecell}
\usepackage{adjustbox}
\usepackage{placeins}
\usepackage[toc,page]{appendix}
\usepackage{fancyhdr}
\usepackage{siunitx}
\ExplSyntaxOn
    \cs_undefine:N \c__siunitx_minus_tl
    \tl_const:Nn \c__siunitx_minus_tl {\scalebox{0.75}[1.0]{$-$}\thinspace}
\ExplSyntaxOff
\graphicspath{{figure_images/}}

\begin{document}

\verso{Jay J. Yoo \textit{et~al.}}

\begin{frontmatter}

\title{Generative adversarial networks for weakly supervised generation and evaluation of brain tumor segmentations on MR images}

\author[1,2,4,6]{Jay J. \snm{Yoo}}
\author[1,2,6]{Khashayar \snm{Namdar}}
\author[2,3]{Matthias W. \snm{Wagner}}
\author[7,8]{Liana \snm{Nobre}}
\author[7,8,9,10]{Uri \snm{Tabori}}
\author[7,9,11,12]{Cynthia \snm{Hawkins}}
\author[2,3]{Birgit B. \snm{Ertl-Wagner}}
\author[1,2,3,4,5,6]{Farzad \snm{Khalvati}\corref{cor1}}
\cortext[cor1]{Corresponding author: 
  \ead{farzad.khalvati@utoronto.ca}}

\address[1]{Institute of Medical Science, University of Toronto, Toronto M5S 1A8, CA}
\address[2]{Department of Diagnostic Imaging, Research Institute, The Hospital for Sick Children, Toronto M5G 1X8, CA}
\address[3]{Department of Medical Imaging, University of Toronto, Toronto M5S 1A8, CA}
\address[4]{Department of Computer Science, University of Toronto, Toronto M5S 1A8, CA}
\address[5]{Department of Mechanical and Industrial Engineering, University of Toronto, Toronto M5S 1A8, CA}
\address[6]{Vector Institute, Toronto M5G 1M1, CA}
\address[7]{The Arthur and Sonia Labatt Brain Tumor Research Centre, The Hospital for Sick Children, Toronto M5G 1X8, CA}
\address[8]{Division of Hematology/Oncology, The Hospital for Sick Children, Toronto M5G 1X8, CA}
\address[9]{Developmental and Stem Cell Biology Program, The Hospital for Sick Children, Toronto M5G 1X8, CA}
\address[10]{Department of Medical Biophysics, University of Toronto, Toronto M5S 1A8, CA}
\address[11]{Department of Laboratory Medicine and Pathobiology, Toronto M5S 1A8, CA}
\address[12]{Department of Paediatric Laboratory Medicine, The Hospital for Sick Children, Toronto M5G 1X8, CA}

\received{} 
\finalform{} 
\accepted{} 
\availableonline{} 
\communicated{} 

\begin{abstract}
Segmentation of regions of interest (ROIs) for identifying abnormalities is a leading problem in medical imaging. Using machine learning for this problem generally requires manually annotated ground-truth segmentations, demanding extensive time and resources from radiologists. This work presents a weakly supervised approach that utilizes binary image-level labels, which are much simpler to acquire, to effectively segment anomalies in 2D magnetic resonance images without ground truth annotations. We train a generative adversarial network (GAN) that converts cancerous images to healthy variants, which are used along with localization seeds as priors to generate improved weakly supervised segmentations. The non-cancerous variants can also be used to evaluate the segmentations in a weakly supervised fashion, which allows for the most effective segmentations to be identified and then applied to downstream clinical classification tasks. On the Multimodal Brain Tumor Segmentation (BraTS) 2020 dataset, our proposed method generates and identifies segmentations that achieve test Dice coefficients of 83.91\%. Using these segmentations for pathology classification results with a test AUC of 93.32\% which is comparable to the test AUC of 95.80\% achieved when using true segmentations. 
\end{abstract}

\begin{keyword}
\KWD Weakly supervised deep learning\sep Low-grade glioma\sep High-grade glioma\sep Image segmentation\sep Generative adversarial networks\sep Magnetic resonance imaging\sep Biomedical imaging
\end{keyword}

\end{frontmatter}


\section{Introduction}
\label{sec:introduction}

Segmenting anomalies in medical images is a leading problem as it can be used to assist in diagnosis and informed patient treatment. However, manually segmenting anomalies demands extensive time and resources from radiologists. Training Machine Learning (ML) models for this segmentation task in a supervised manner requires ground-truth manual segmentations that are also cumbersome to acquire \cite{MCDONALD20151191, Razzak2018}. Furthermore, segmentation models often demand large datasets of annotated medical images, which may not be available for specific contexts such as pediatric cancer. Hence, training ML models with limited or without ground truth segmentation masks is key. 

Weakly supervised segmentation simplifies the required ground truth annotation. This simplification can be done through various means such as using partial annotations \cite{Liang_Nan_Coppola_Zou_Sun_Zhang_Wang_Yu_2019}, bounding boxes \cite{Yang_Wang_Yang_Chen_Tang_Shao_Dillenseger_Shu_Luo_2020, Zhang_Chen_Chong_Li_2021}, circular centroid labels \cite{Li_Wang_Liu_Latecki_Wang_Huang_2019}, or scribble marks \cite{10.1007/978-3-030-32248-9_20}. We opt to use the simplest form of weakly supervised labels; binary image-level labels indicating whether the image has the anomaly of interest or not. 

The common approach to weakly supervised segmentation using binary image-level labels is to train a classification model and then using the classification model to infer tumor segmentations. The most popular means of inferring tumor segmentations from classification models is to use class activation maps (CAMs) \cite{10.1007/978-3-319-66179-7_65,10.1007/978-3-030-32248-9_24,tang2020automated,7780688,PATEL2022102374}. There have been other approaches, such as classifying patches of an image to form rough segmentations \cite{lerousseau2021weakly} or generating explanation maps using methods including Local Interpretable Model-Agnostic Explanations (LIME) \cite{10.1145/2939672.2939778}, SHapley Additive exPlanations (SHAP) \cite{Lundberg_Erion_Chen_DeGrave_Prutkin_Nair_Katz_Himmelfarb_Bansal_Lee_2020}, or Randomized Input Sampling for Explanation of Black-box Models (RISE) \cite{Petsiuk2018RISERI}. However, a more recent approach to weakly supervised segmentation is to use Generative Adversarial Networks (GANs) \cite{10.1145/3422622}. Han et al. \cite{8869751} demonstrated that GANs could generate realistic brain Magnetic Resonance (MR) images. Their work used a standard GAN approach using a vector from a latent Gaussian space as input to generate brain MR images, demonstrating the potential for realistic brain MR images to be produced from a Gaussian distribution. This idea has been explored for weakly supervised segmentation by generating a healthy version of cancerous brains and then subtracting it from the original image to acquire segmentations \cite{10.1007/978-3-030-11723-8_16,Baur_Wiestler_Muehlau_Zimmer_Navab_Albarqouni_2021}. 

We propose a new methodology for weakly supervised segmentation on 2-dimensional (2D) MR images that only requires binary image-level labels which indicate the presence of glioma to generate effective glioma segmentations. To do so, we use the image-level labels to generate two priors of information. The first is localization seeds acquired from a binary classifier that indicate which regions are likely to contain a tumor and which regions are least likely to contain a tumor. The second is from non-cancerous variants of cancerous MR images that we generate using GANs. These priors used in conjunction are more effective than using either prior independently for generating segmentations. We choose to explore 2D segmentation because many institutional datasets lack the size required to train GANs or weakly supervised segmentation models. Converting 3D patient volumes to 2D slices artificially increases dataset size at the cost of 3D context. 

The non-cancerous variants can also be used to evaluate the generated weakly supervised segmentations, identifying the most effective segmentations from each patient volume. Radiomic features from these segmentations can then be used to train models for downstream glioma clinical classification tasks. This enables the 2D segmentations to be applied to 3D clinical classification.

The main contributions of the work are summarized as follows:
\begin{itemize}
    \item We present a method for generating effective 2D weakly supervised segmentations of brain tumors on MR images by using non-cancerous variants of cancerous MR brain images generated from GANs.
    \item We propose a metric for estimating the quality of weakly supervised segmentations in a weakly supervised manner using the generated non-cancerous variants.
    \item We demonstrate that using weakly supervised segmentations generated using our method with the proposed metric enables effective 3D brain tumor classification. 
\end{itemize}

\section{Methods}

\begin{figure}[!htbp]
  \centering
  \includegraphics[width=\linewidth]{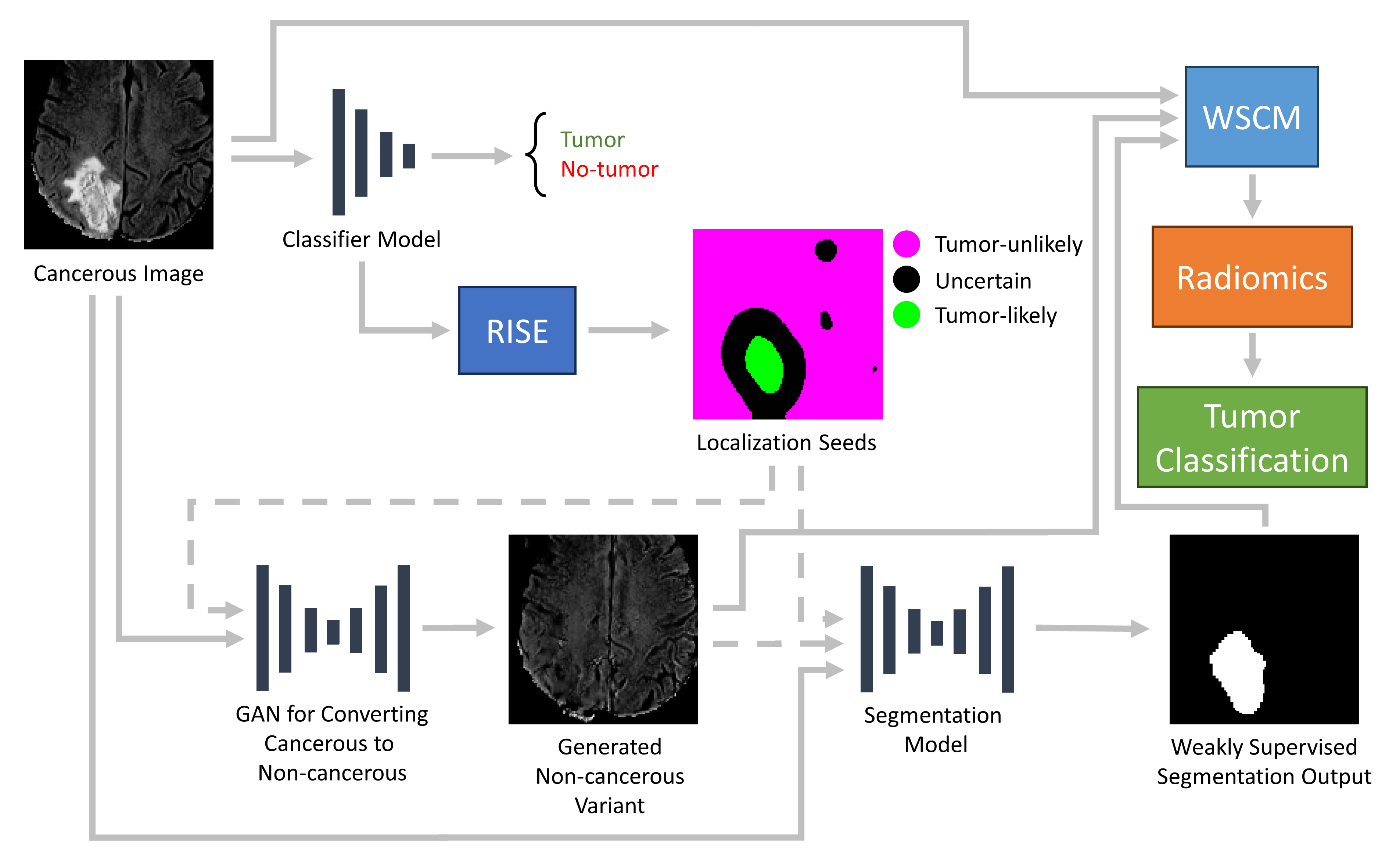}
  \caption{Block diagram visualizing the overall method. Solid gray lines indicate direct inputs and outputs. Dashed gray lines indicate use in loss functions. }
  \label{fig:holistic_diagram}
\end{figure}

Figure \ref{fig:holistic_diagram} visualizes different steps for our proposed weakly supervised segmentation method, and how the outputs from each step are used in the overall methodology. 

\subsection{Seed Generation}

We first generate seeds to inform which regions in the images are likely to contain tumors and which regions are unlikely to contain any tumors. These seeds will be used as the primary source of localization information when training the weakly supervised segmentation models. 

We define the 2D MR images as $X = \{x^0, ..., x^k, ..., x^{N - 1}\}$, $x^k \in \mathbb{R}^{c, 128, 128}$, where $N$ is the number of images and $c$ is the number of channels. We also define the binary-image level labels as $Y = \{y^0, ..., y^k, ..., y^{N - 1}\}$, $y^k \in \mathbb{R}$ where each slice $y^k$ was assigned $0$ if the corresponding image $x^k$ was not cancerous, and $1$ otherwise. To generate the seeds, we upsample the spatial dimensions of $X$ from $128 \times 128$ to $256 \times 256$ using bilinear interpolation and then use the upsampled data with the binary-image level labels to train a classifier model $C()$. The predicted classification scores for each image in $X$ are defined as $P = \{p^0, ..., p^k, ..., p^{N-1}\} \in \mathbb{R} \mid 0 \leq p^k \leq 1$, where $p^k > 0.5$ indicates that $C()$ predicts $x^k$ to be cancerous.

We used the RISE method \cite{Petsiuk2018RISERI} and normalized the outputs to range $[0, 1]$ to acquire the explanation maps $E = \{{e^0, ..., e^k, ..., e^{N-1}}\} \in \mathbb{R}^{N, H, W}$ from $C()$ as was done by Yoo et al. \cite{yoo2024deep}.

\begin{figure}[!htbp]
  \centering
  \includegraphics[width=0.45\linewidth]{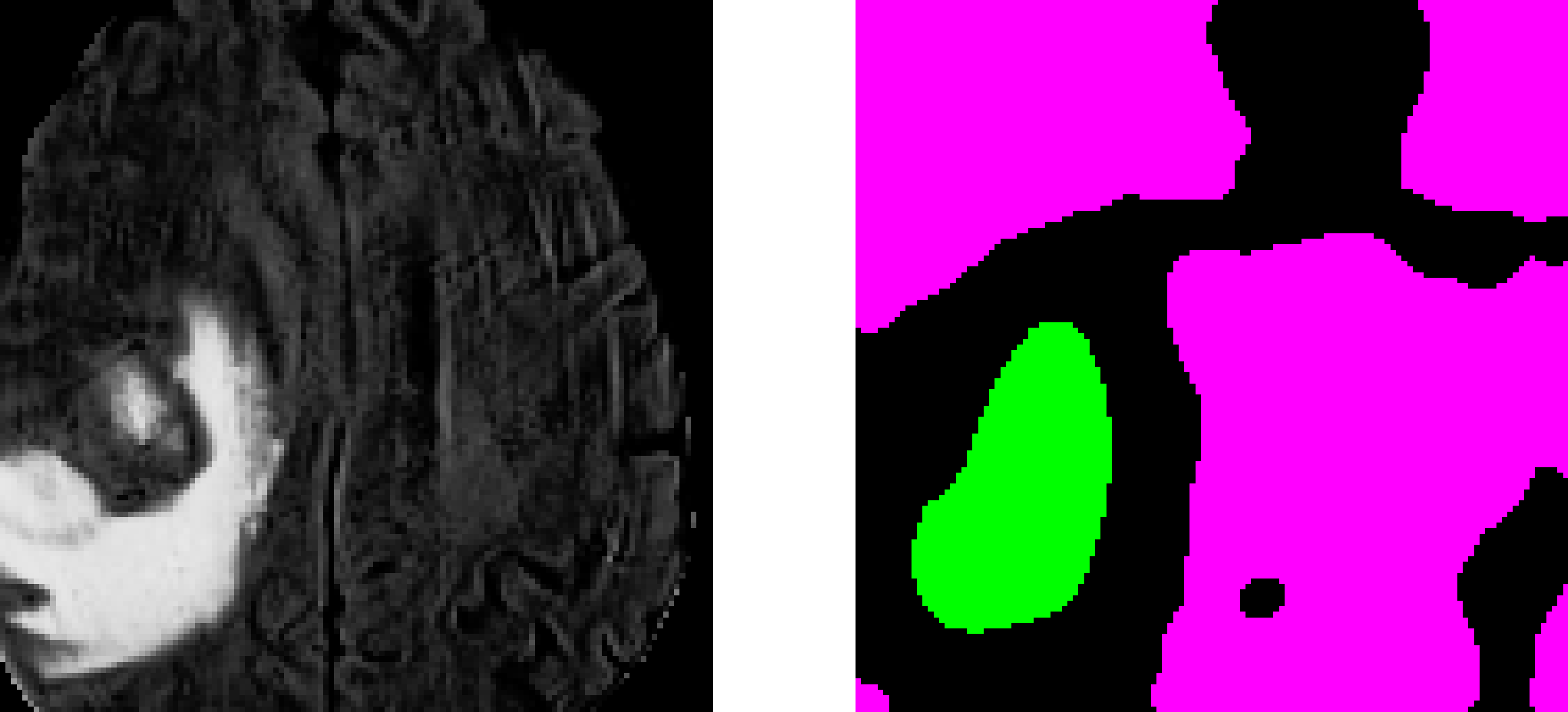}
  \caption{MR image FLAIR channel (left) and its corresponding positive seeds in green, negative seeds in magenta, and unseeded regions in black (right).}
  \label{fig:initial_seeds}
\end{figure}

Once $E$ is generated, we use it to acquire positive seed regions and negative seed regions. Positive seed regions $S_+ = \{{s_+^0, ..., s_+^k, ...., s_+^{N-1}}\} \in \mathbb{R}^{N, 128, 128}$ are binary maps where for each $s_+^k$, a voxel $(i, j)$ has a value of 1 when the corresponding voxel in $e^k$ has a value greater than 0.8 and 0 otherwise. The voxels in the negative seed regions $S_- = \{{s_-^0, ..., s_-^k, ...., s_-^{N-1}}\} \in \mathbb{R}^{N, 128, 128}$ have values of 1 for the corresponding voxels in $E$ that have values less than 0.2 and 0 otherwise. The positive seeds are used to indicate regions in the images that $C()$ is confident contains a tumor and negative seeds are used to indicate regions $C()$ considered to contain no tumor. The 20\% threshold we used to generate the seeds was suggested by \cite{7780688} in their work with segmentation using CAMs. For non-cancerous MR images, we used a positive seed map of all zeros and a negative seed map of all ones. Figure \ref{fig:initial_seeds} presents examples of positive and negatives seeds. 

\subsection{Generation of Non-cancerous Variations of Cancerous MR Images}

Once seeds have been generated, we train a model that receives cancerous images as input and outputs non-cancerous variations of the input images. The non-cancerous variants can be compared to their original to further inform the training of the desired weakly supervised segmentation model. To generate the non-cancerous variants we use a U-Net model that consists of an encoder and generator connected by skip connections \cite{10.1007/978-3-319-24574-4_28}. The generator is trained individually first to initialize the weights and guide the training toward the desired non-cancerous image outputs.

\subsubsection{Generator Training}
\label{sec:generator_training}

To generate non-cancerous variants of cancerous MR images, we first define subsets of $X$ from the training cohort, $X_C$ contains all images that $C()$ correctly believes to be cancerous and $X_{NC}$ contains all images that are non-cancerous. We exclude images with false positive predictions from $C()$ because these images are assumed to have poor seeds. The T2-FLAIR channel of $X_{NC}$, defined as $X_{NC,FLAIR}$ is then used to train a generator model $G()$ in a deep convolutional GAN (DCGAN) system that maps the space of unit Gaussian latent vectors to the space of non-cancerous MR images $X_{NC,FLAIR}$. Solely using the T2-FLAIR channel for training the DCGAN was also done by Baur et al. \cite{Baur_Wiestler_Muehlau_Zimmer_Navab_Albarqouni_2021} and serves to decreases the complexity of the learning task.

A latent vector $z$ sampled from the unit Gaussian space is passed into $G()$ which then outputs a generated image $g$. $g$ is then passed into a discriminator model $D_0()$, which is trained to distinguish real and generated images using the binary cross-entropy (BCE) loss function, defined as $L_{D_0}$ in Equation \eqref{eq:d0_loss}. $D_0()$ is trained in two batches, each of which is of the same size $B$. The first batch consists of real images from $X_{NC,FLAIR}$ and the second batch consists of the generated images. For training the generator, we used one-sided label smoothing by setting the image labels to $0.9$ for the batch of real images and $0.1$ for the batch of generated images \cite{7780677}, which prevents the discriminator from becoming too confident in its classifications thereby allowing the generator to train effectively. $L_{D_0}$ is then used to calculate the loss function of $G()$, $L_G$, such that $G()$ learns to generate images that $D_0()$ discerns as real non-cancerous brain MR images. This system is expressed mathematically in equations \ref{eq:d0_loss}-\ref{eq:g_loss}. In these equations, the losses are computed over all N but it should be noted that in implementation, minibatches were used, and thus rather than computing the losses over N, they are computed over each batch. For simplicity, the loss equations will be presented as being computed over N.

\begin{equation}
    \label{eq:d0_loss}
    \begin{aligned}
    L_{D_0} &= - \frac{1}{N} \sum_{k=0}^{N - 1} \left( y_0^k \log (D_0(x_\text{in}^k)) + \right. \\ 
    &\left. (1 - y_0^k) \log (1 - D_0(x_\text{in}^k)) \right), \\
    y_0^k &= \begin{cases}
      0.1, & \text{if}\ x_\text{in}^k = g \\
      0.9, & \text{if}\ x_\text{in}^k = x_{NC,FLAIR}^k
    \end{cases}
    \end{aligned}
\end{equation}

\begin{equation}
    \label{eq:generated_image}
    g = G(z), z \sim N(0, 1)
\end{equation}

\begin{equation}
    \label{eq:g_loss}
    L_G = - \frac{1}{N} \sum_{k=0}^{N - 1} \log (D_0(g))
\end{equation}

\subsubsection{U-Net Training Initialized with Generator Weights from Generator Training}

\begin{figure}[!htbp]
  \centering
  \includegraphics[width=
  \linewidth]{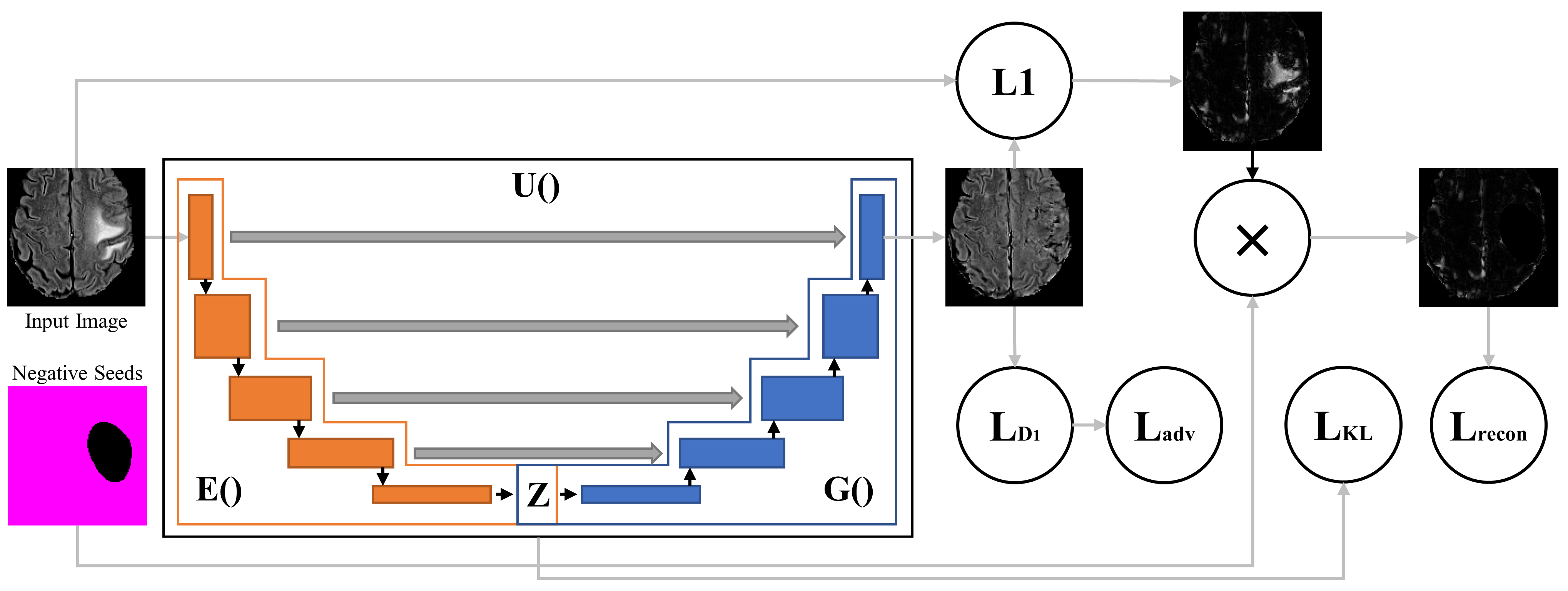}
  \caption{Architectures for generator $G()$, encoder $E()$, and converting U-Net model $U()$, as well as the calculation of the reconstruction loss $L_{recon}$.}
  \label{fig:normalizing_flow_architecture}
\end{figure}

Once $G()$ is trained, it is appended by an encoding model $E()$ using skip connections to form a U-Net model, denoted as $U()$. This U-Net model is then trained in a similar paradigm described in Section \ref{sec:generator_training} with a few differences. The generated image is no longer acquired from passing a unit Gaussian latent vector to G. Instead, a cancerous image $x_{C,FLAIR}^k$ is passed into the U-Net model and the model outputs a non-cancerous version of the image $\hat{x}_{C,FLAIR}^k$, which is used as the generated image in the DCGAN system. The architecture of $G()$, $E()$, and $U()$ is visualized in Figure \ref{fig:normalizing_flow_architecture}. The U-Net model is trained using three loss terms. The first is the adversarial loss $L_{adv}$ which, like $L_G$, uses a BCE loss to reward $U()$ for outputting images that a new discriminator $D_1()$, trained using $L_{D_1}$, classifies as being non-cancerous. The second loss is the reconstruction loss $L_{recon}$ that rewards the U-Net model for generating images that matches the input image at negative seed regions. It does so by measuring the absolute difference between the cancerous image and its non-cancerous variant in the negative seed regions. This has the effect of encouraging the model to keep non-cancerous regions of the cancerous images unchanged. The third loss is a Kullback-Leibler (KL) divergence loss \cite{10.1214/aoms/1177729694}, $L_{KL}$ that is applied to the encodings outputted by the final layer of $E()$. This loss encourages the outputs of $E()$, $\hat{z}$ to be within the space of unit Gaussian vectors. Not only does this have a regularizing effect on $U()$ but it also enables $G()$ to be initialized using the weights acquired from the previous training as those weights were trained using unit Gaussian inputs to $G()$. These loss terms are described in Equations \ref{eq:encoder_output}-\ref{eq:normalizingflow_loss}.

\begin{equation}
    \label{eq:encoder_output}
    \begin{aligned}
    \hat{z}^k = E(x_{C,FLAIR}^k)
    \end{aligned}
\end{equation}

\begin{equation}
    \label{eq:normalizing_flow_output}
    \begin{aligned}
    \hat{x}^k_{C,FLAIR} = U(x_{C,FLAIR}^k)
    \end{aligned}
\end{equation}

\begin{equation}
    \label{eq:d1_loss}
    \begin{aligned}
    L_{D_1} &= - \frac{1}{N} \sum_{k=0}^{N - 1} \left( y_1^k \log (D_1(x_\text{in}^k)) \right. \\
    &\left. + (1 - y_1^k) \log (1 - D_1(x_\text{in}^k)) \right), \\
    y_1^k &= \begin{cases}
      0.1, & \text{if}\ x_\text{in}^k = \hat{x}_{C,FLAIR}^k \\
      0.9, & \text{if}\ x_\text{in}^k = x_{NC,FLAIR}^k
    \end{cases}
    \end{aligned}
\end{equation}

\begin{equation}
    \label{eq:adv_loss}
    \begin{aligned}
    L_{adv} =& - \frac{1}{N} \sum_{k=0}^{N - 1} \log (D_1(\hat{x}_{C,FLAIR}^k))
    \end{aligned}
\end{equation}

\begin{equation}
    \label{eq:recon_loss}
    \begin{aligned}
    L_{recon} =& \frac{1}{N} \sum_{k=0}^{N - 1} \left( \frac{1}{\sum_{i,j}^{H,W} s_-^k} \sum_{i,j}^{H,W} {s_-^k}_{i,j} \lvert x_{C,FLAIR}^k \right. \\
    & \Bigg. - \hat{x}_{C,FLAIR}^k \rvert_{i,j} \Bigg)
    \end{aligned}
\end{equation}

\begin{equation}
    \label{eq:kl_loss}
    \begin{aligned}
    L_{KL} =& - \frac{1}{N} \sum_{k=0}^{N - 1} \left( \left( 2 \log (\sigma_{\hat{z}^k}) \right) -\sigma_{\hat{z}^k}^2 - \mu_{\hat{z}^k}^2 + 1 \right)
    \end{aligned}
\end{equation}

\begin{equation}
    \label{eq:normalizingflow_loss}
    \begin{aligned}
    L_{U} = \alpha L_{recon} + \beta L_{adv} + \gamma L_{KL}
    \end{aligned}
\end{equation}

Note that $\mu_{\hat{z}^k}$ and $\sigma_{\hat{z}^k}$ are the mean and standard deviation of $\hat{z}^k$, respectively. In Equation \eqref{eq:normalizingflow_loss}, $\alpha$, $\beta$, and $\gamma$ are weights used to balance the loss terms. We refer to $\alpha$ as the reconstruction scale, $\beta$ as the adversarial scale, and $\gamma$ as the KL scale. When setting the reconstruction scale and adversarial scale to similar values, the adversarial loss tends to dominate the reconstruction loss, resulting with the model tending to generate non-cancerous images that are completely different from the input cancerous images. Thus, the reconstruction scale is required to be significantly greater than the adversarial scale for the model to consistently output non-cancerous variations of the input cancerous images. 

Rather than training $U()$ with the whole $X_{C,FLAIR}$, we use a subset of $X_{C,FLAIR}$ whose negative seed regions consist of less than $50\%$ of the image. This is to ensure that when the negative seed region is applied to the input image, a sufficient amount of the image remains for $L_{recon}$. 

We trained $G()$ prior to training $U()$ instead of training $U()$ from scratch because we found that it can be difficult for the U-Net model to learn to output the desired non-cancerous variants without pretraining $G()$.

\subsection{Segmentation}
\label{sec:initial_segmentation}

To generate segmentations using the generated non-cancerous variations $\hat{X}_{C,FLAIR}$, we use the same U-Net architecture as $U()$. This new U-Net model, denoted as $S()$, is trained using three loss terms and all the available data $X$ to generate segmentations $M = \{m, ..., m^k, ..., m^{N-1}\}$. The first loss term is $L_{seed}$ defined in Equation \eqref{eq:seed_loss_2}. In this case, we use the positive seeds $S_+$ and negative seeds $S_-$. This loss, based on the seeding loss proposed by Kolesnikov and Lampert \cite{kolesnikov2016seed}, encourages the model generate segmentations that include the positive seeds and exclude the negative seeds by measuring the total value of the log pixel intensities in the generated segmentation corresponding to the positive seeds and the negative seeds. The second loss term is the variation reconstruction loss $L_{var}$ which attempts to minimize the L1 loss between an image $x^k$ and its non-cancerous variant $\hat{x}^k$ in regions not segmented by $m^k$. This encourages the model to segment regions of cancerous images that differ greatly from their non-cancerous counterparts, i.e. tumor locations. For non-cancerous images $X_{NC}$, their non-cancerous variants are considered identical, that is, $X_{NC,FLAIR} = \hat{X}_{NC,FLAIR}$. To clarify, the segmentation model is trained on all available channels, unlike the GAN. However, computing $L_{var}$ only uses the FLAIR channel since it uses outputs from the GAN which was trained only on the FLAIR channel. Since the output segmentations cannot be converted to binary segmentations in a differentiable manner, we multiply the voxel-wise L1 loss map between $x_{FLAIR}^k$ and $\hat{x}_{FLAIR}^k$ by $1 - m^k$, as the seed loss will encourage the model's outputs to be close to 0 or 1 due to $S_+$ and $S_-$ being binary. Although non-cancerous variants are not sufficient for segmentation on their own, $L_{var}$ enables them to inform the segmentation model of the shape of the tumors. The final loss term is a size loss $L_{size}$ that simply encourages the model to minimize the sum of its outputs. As the seed loss encourages outputs close to 0 or 1, the size loss encourages the output segmentations to have tight boundaries around the tumors, as this is the only way to minimize the size loss without compromising the other loss terms. This system is visualized in Figure \ref{fig:segmentation_model_system} and these loss terms are expressed in Equations \ref{eq:seg0_output}-\ref{eq:seg0_loss}.

\begin{figure}[!htbp]
  \centering
  \includegraphics[width=0.8\linewidth]{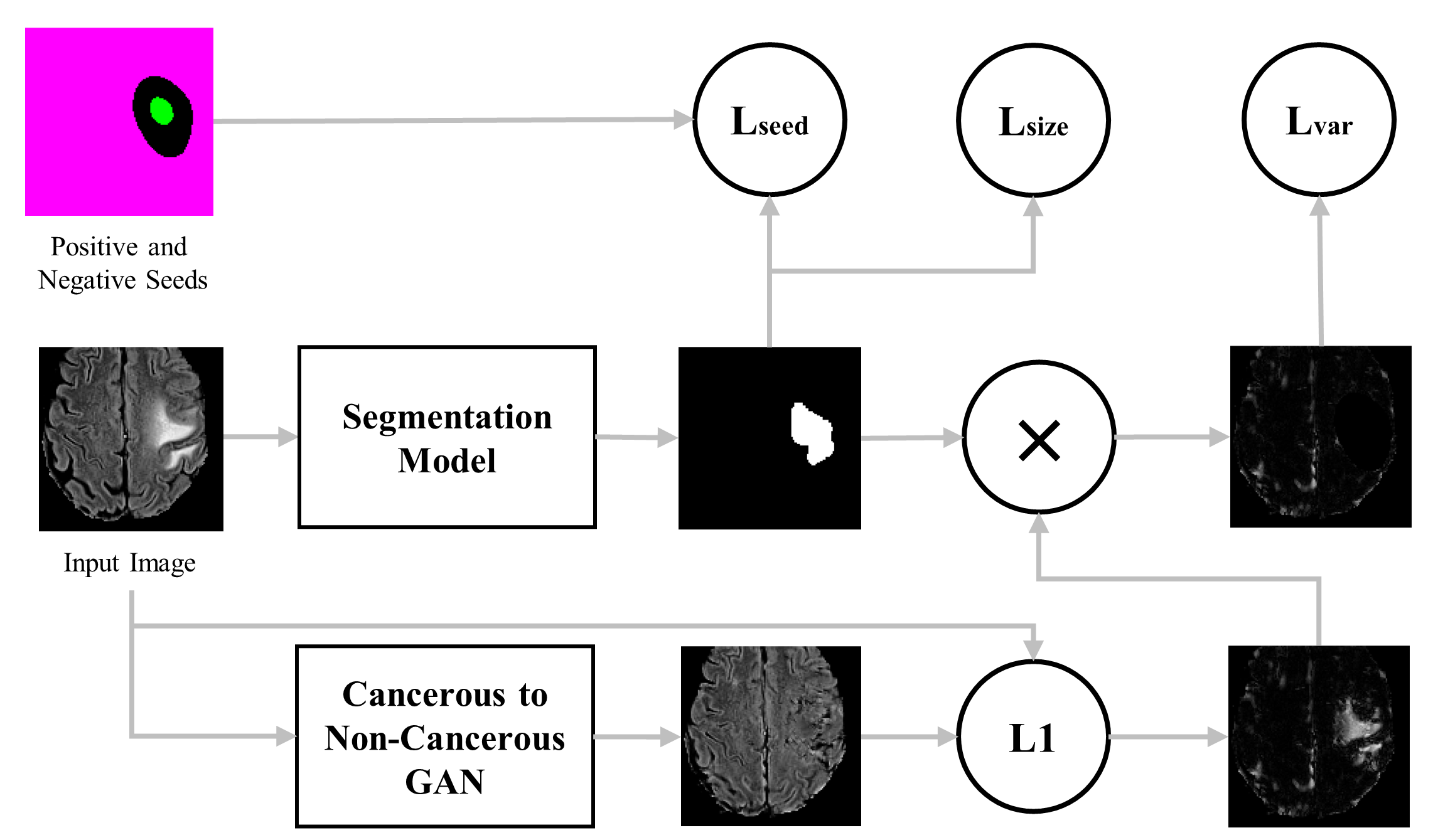}
  \caption{System for training the segmentation model using the seed loss, variation reconstruction loss, and size loss.}
  \label{fig:segmentation_model_system}
\end{figure}

\begin{equation}
    \label{eq:seg0_output}
    \begin{aligned}
    m^k &= S(x^k)
    \end{aligned}
\end{equation}

\begin{equation}
    \begin{aligned}
    \label{eq:seed_loss_2}
    L_{seed} &= \frac{1}{N} \sum_{k=0}^{N - 1} \left( \frac{-1}{\sum_{C \in [+, -]} \lvert {s_C^k} \rvert} \right. \\
    & \Bigg. \sum_{C \in [+, -]} \sum_{i, j \in s^k_C} \log \left( {m^k}_{i,j} \right) \Bigg)
    \end{aligned}
\end{equation}

\begin{equation}
    \label{eq:recon_loss_2}
    \begin{aligned}
    L_{var} =& \frac{1}{N} \sum_{k=0}^{N - 1} \left( \frac{1}{\sum_{i,j}^{H,W} s_-^k} \sum_{i,j}^{H,W} {s_-^k}_{i,j} \lvert x_{FLAIR}^k \right. \\
    & \Bigg. - \hat{x}_{FLAIR}^k \rvert_{i,j} \Big)
    \end{aligned}
\end{equation}

\begin{equation}
    \label{eq:size_loss}
    \begin{aligned}
    L_{size} =& \frac{1}{N} \sum_{k=0}^{N - 1} \left( \frac{1}{HW} \sum_{i,j}^{H,W} {m^k}_{i,j} \right)
    \end{aligned}
\end{equation}

\begin{equation}
    \label{eq:seg0_loss}
    \begin{aligned}
    L_{M} = \delta L_{seed} + \epsilon L_{var} + \zeta L_{size}
    \end{aligned}
\end{equation}

$\delta$, $\epsilon$, and $\zeta$ from Equation \eqref{eq:seg0_loss} are weights to balance the loss terms for $L_{M}$. We set these weights to $\delta = 4$, $\epsilon = 1$, and $\zeta = 0.25$. The weight for $L_{size}$ is set to the lowest value between the three weights because we do not want the size loss to overpower the other losses and begin excessively shrinking the segmentations or decreasing the intensity values of the segmented voxels. Since the non-cancerous variations do not always differ from their original counterparts solely at the tumor locations, we rely on the seeds more than the generated images. Thus, the weight for $L_{seed}$ is greater than the weight for $L_{var}$. 

When acquiring segmentations from $S()$, for evaluation, we assume that the images predicted by $C()$ to have no tumor have segmentation maps consisting of all zero values. The output segmentation is defined as $m^k$

\subsection{Weakly Supervised Confidence Measure of Segmentations}
\label{sec:method_radiomics}
We propose an approximate confidence measure of the quality of the segmentations that can be computed without using the true segmentations, which we refer to as weakly supervised confidence measure (WSCM). The WSCM $c^k$ for each segmentation $m^k$ is measured by calculating the mean absolute difference between an image $x^k$ and its non-cancerous variant $\hat{x}^k$ in the segmentation region defined by $m^k$, as expressed in Equation \eqref{eq:wscm}. 

\begin{equation}
    \label{eq:wscm}
    c^k = \begin{cases}
      \frac{1}{\mid m^k \mid} \sum_{i, j \in m^k} \mid x^k_{FLAIR_{i,j}} - \\
      \hat{x}_{FLAIR_{i, j}}^k \mid , & \text{if}\ \mid m^k \mid > 0 \\
      1, & \text{otherwise}\
    \end{cases}
\end{equation}

\subsection{Radiomics-based Classification using Weakly Supervised Segmentations}
One use of tumor segmentations for downstream clinical tasks is to extract Radiomic features from the segmentations and then classify the tumors using the extracted features. The common approach for using 2D segmentations for 3D classification is to select the axial slice with the largest 2D segmentation for each 3D volume. However, this approach to slice selection cannot be applied to weakly supervised segmentations because there is a risk that oversegmented slices are selected. We instead inform the slice selection using our proposed WSCM metric. To do so, we first identified the top 25\% axial segmentations according to WSCM for each volume. We then selected the 5 largest segmentations amongst the identified segmentations, resulting with 5 axial segmentations for each volume. 

PyRadiomics \cite{pyradiomics} was then used to extract the Radiomic features from the segmentations of the 5 selected axial slices for each patient volume. Due to PyRadiomics only being able to extract features from single channel inputs, each channel had their own Radiomic features extracted for multimodal MR image inputs before being concatenated together. The diagnostic and shape features were removed. Features with correlation greater than 0.95 or variance less than 0.05 were also removed. The remaining features were then min-max normalized to range [0, 1]. Then, the training cohort was augmented using the Synthetic Minority Oversampling TEchnique \cite{smote}. The preprocessed features were then used to train random forest models to predict the pathology or genetic marker of the tumors. The Radiomic features extracted from each axial slice were treated as individual samples during training. During evaluation, the predicted probabilities from the 5 axial slices for each volume were averaged together to form a single probability score for each volume. 

\section{Datasets}
\label{sec:datasets}

We evaluated our proposed weakly supervised on two brain MR imaging datasets. The first is the open source Multimodal Brain Tumor Segmentation (BraTS) Challenge 2020 dataset \cite{menze_multimodal_2015}--\cite{bakas_segmentation_2017-1} and the second is an internal dataset of pediatric low-grade neuroepithelial tumors (PLGNT). For the BraTS dataset, we preprocessed the images to spatial dimensions of $128 \times 128$ for training but to evaluate the performance of the trained models, we segmented the whole 2D MR images. To do so, we split the whole 2D MR images into patches with spatial dimensions of $128 \times 128$. The segmentation model is then applied to each patch. The segmentation probability maps for all patches are then reconstructed to a whole image segmentation probability map, with the probability maps being averaged at regions with overlapping patches, before being binarized by setting all values greater than $0.5$ to $1$ and all other values to $0$. 

\subsection{Multimodal Brain Tumor Segmentation 2020 Dataset}
To form the primary target dataset for this study, the native (T1-weighted), post-contrast T1-weighted, T2-weighted, and  T2 Fluid Attenuated Inversion Recovery (T2-FLAIR) volumes from the open source BraTS 2020 dataset containing high-grade glioma (HGG) or low-grade glioma (LGG) were first randomly divided into training, validation and test cohorts using 80/10/10 splits. Each volume was then preprocessed by first cropping each image and segmentation map using the smallest bounding box which contained the brain, clipping all non-zero intensity values to the 1 and 99 percentiles, normalizing the cropped volumes using min-max scaling, and then randomly cropping and padding the volumes to fixed patches of size $128 \times 128$ along the coronal and sagittal axes, as done by Henry et al. \cite{henry_brain_2020} in their work with BraTS datasets. The volumes and segmentation maps were then split into axial slices to form sets of stacked 2D images with 4 channels $x^k \in \mathbb{R}^{4, H, W}$, forming the set $X = \{x^0, ..., x^k, ..., x^{N - 1}\}$, where $N$ is the number of images in the dataset, and both $H$ and $W$ are $128$ when training. Slices with more than 50\% of voxels equal to zero were removed from $X$. Only the training set of the BraTS dataset was used because it is the only one with publicly available ground truths. 

The dataset offers ground-truth segmentations which have to be converted to binary labels to match the weakly supervised task. The segmentations are provided as values that are 0 where there are no tumors and an integer between 1-4 depending on the region of the tumor in which a voxel is located. As this work focuses solely on segmenting the whole tumor, a new set of ground truths $Y = \{y^0, ..., y^k, ..., y^{N - 1}\}$, $y^k \in \mathbb{R}$ is defined, where for each slice $y^k$ was assigned $0$ if the corresponding segmentations were empty, and $1$ otherwise. To account for images with tumors that are too small to be effectively segmented, small labels were removed using morphological erosion. 

369 patient volumes from the BraTS 2020 training set were split into 295, 37, and 37 volumes for the training, validation, and test cohorts, respectively. After splitting the volumes into 2D images, the first 30 and last 30 slices of each volume were removed, as done by Han et al. \cite{8869751} because these slices lack useful information. The training, validation, and test cohorts had 24635, 3095, and 3077 stacked 2D images, respectively, of which 68.93\%, 66.30\%, and 72.23\% of the slices are cancerous. 

\subsection{Pediatric Low-Grade Neuroepithelial Tumors Dataset}
We also evalueted the proposed method using an internal dataset from the Hospital for Sick Children consisting of 340 T2-FLAIR patient volumes containing PLGNT. This retrospective study was approved by the institutional Research Ethics Board (REB) (ethics board protocol number: \#1000077114; approved May 26, 2021). Due to the retrospective nature of the study, informed consent was waived by the institutional REB. This dataset consists of 340 volumes from patients ranging from 0 to 18 years of age who were identified using the electronic health record database of the hospital from January 2000 to December 2018. The volumes are labelled with a variety of genetic markers including BRAF fusion and BRAF V600E mutation. There were two main differences in the preprocessing performed for this secondary target dataset compared to the preprocessing done for the BraTS dataset. The first difference was that instead of cropping, the volumes were downsampled from $240 \times 240$ to $120 \times 120$ and then randomly padded to patches of size $128 \times 128$ along the coronal and sagittal axes. Given that only the FLAIR channel was available, the second difference was that the volumes for this dataset were split into axial slices forming stacked 2D images with 1 channel instead of 4 channels, resulting in the dataset consisting of images $X = \{x^0, ..., x^k, ..., x^{N - 1}\}$, $x^k \in \mathbb{R}^{1, H, W}$, where $N$ is the number of images in the dataset, and both $H$ and $W$ are $128$ when training. 

After preprocessing, the 340 PLGNT patient volumes were split into 272, 34, and 34 volumes for the training, validation, and test cohorts respectively. After splitting into 2D images and removing the first and last 30 slices from each volume, the training, validation and test cohorts had 24499, 3077, and 3151 stacked 2D images respectively, of which 33.48\%, 37.67\%, and 34.53\% of were cancerous. When performing Radiomics-based classification on the PLGNT dataset, we chose to exclude volumes that did not have genetic markers of BRAF fusion or BRAF V600E mutation, to ensure both datasets had a binary classification task. This resulted with 173, 19, and 22 volumes in the training, validation, and test cohorts respectively when performing classification. 

\section{Results}
\label{sec:results}

\subsection{Segmentation Model Performance}

\begin{figure}[!t]
  \centering
  \subfloat[Example 1.]{%
  \includegraphics[width=0.45\linewidth]{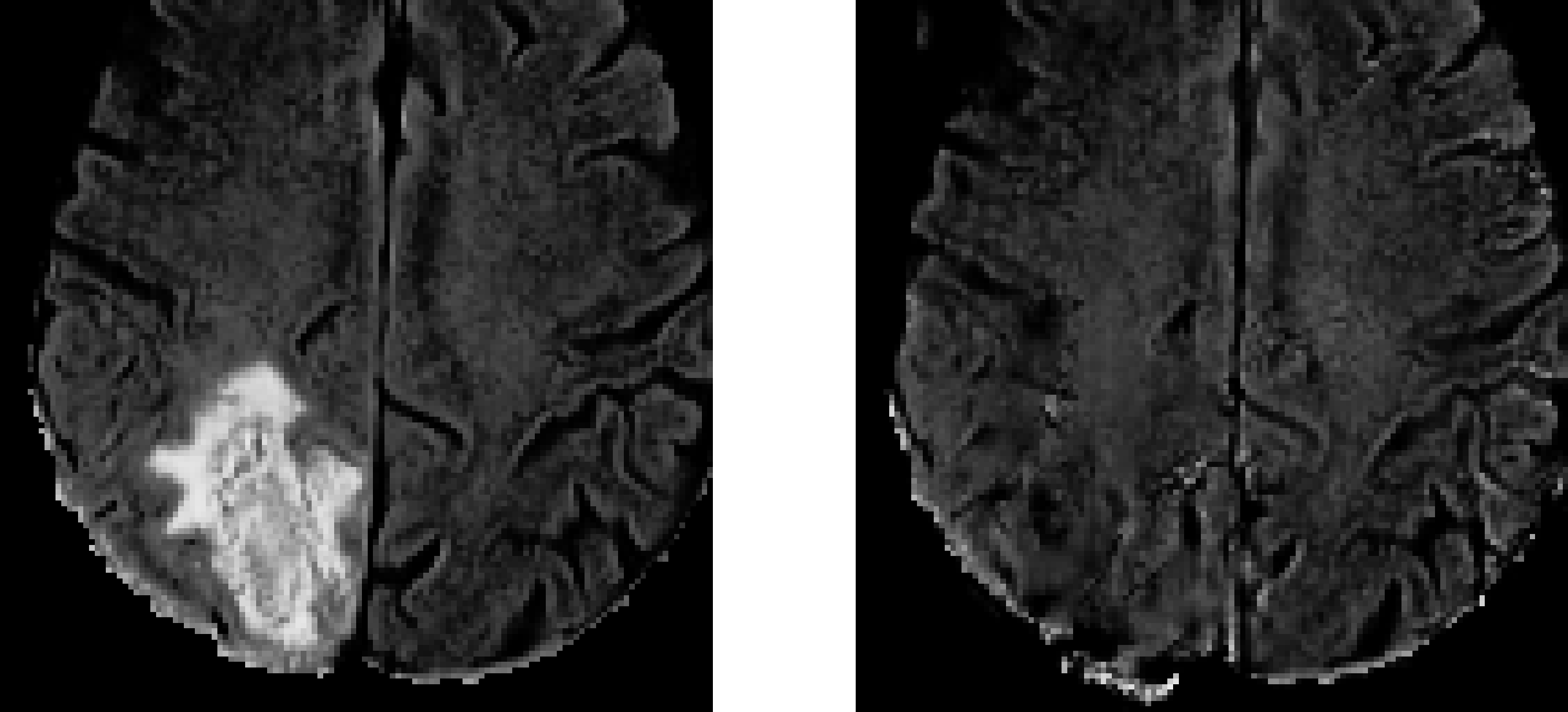}
  }\hfill
  \subfloat[Example 2.]{%
  \includegraphics[width=0.45\linewidth]{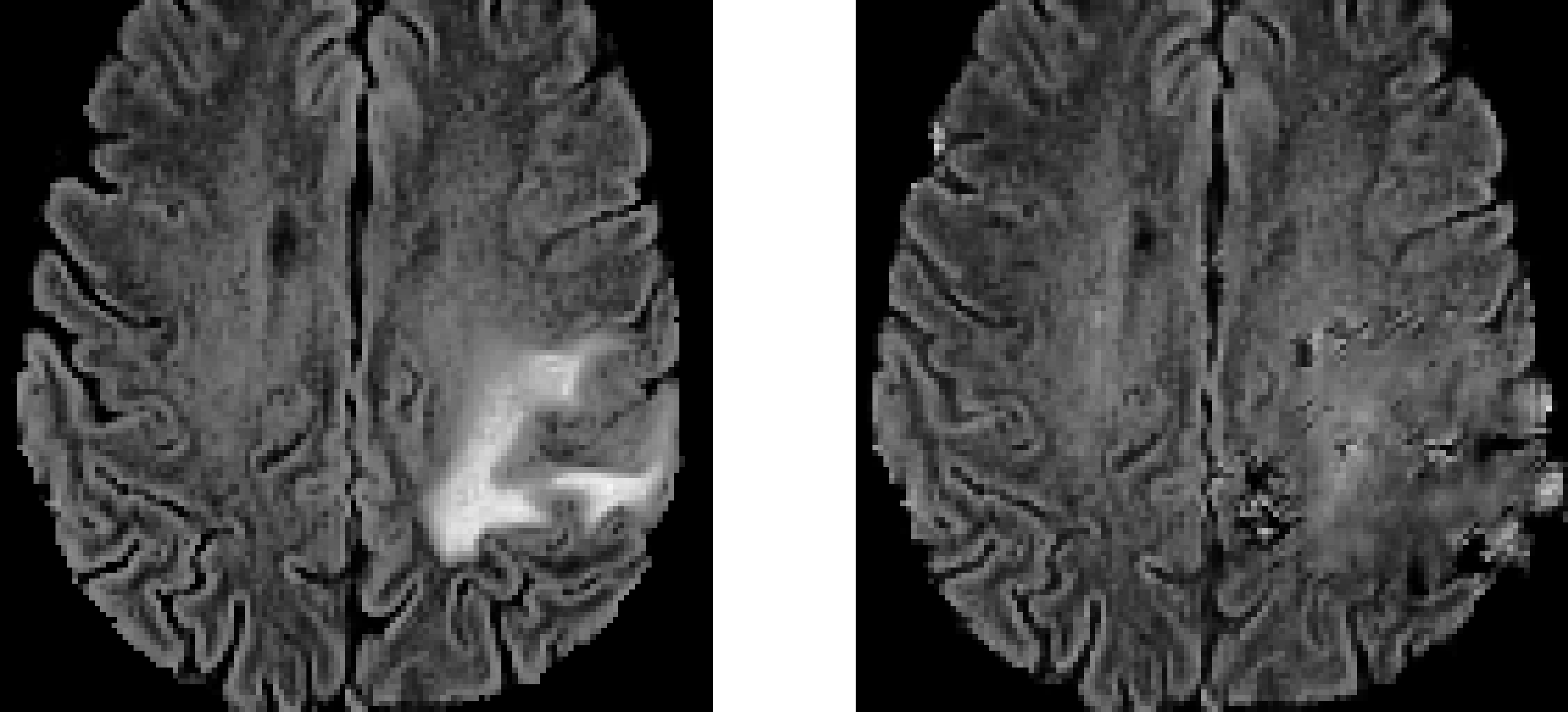}
  }
  \caption{Examples of MR image FLAIR channel (left) and their generated non-cancerous variant (right).}
  \label{fig:conversion}
\end{figure}

\begin{figure}[h!]
  \centering
  \subfloat[Examples from BraTS dataset.]{%
  \includegraphics[width=0.45\linewidth]{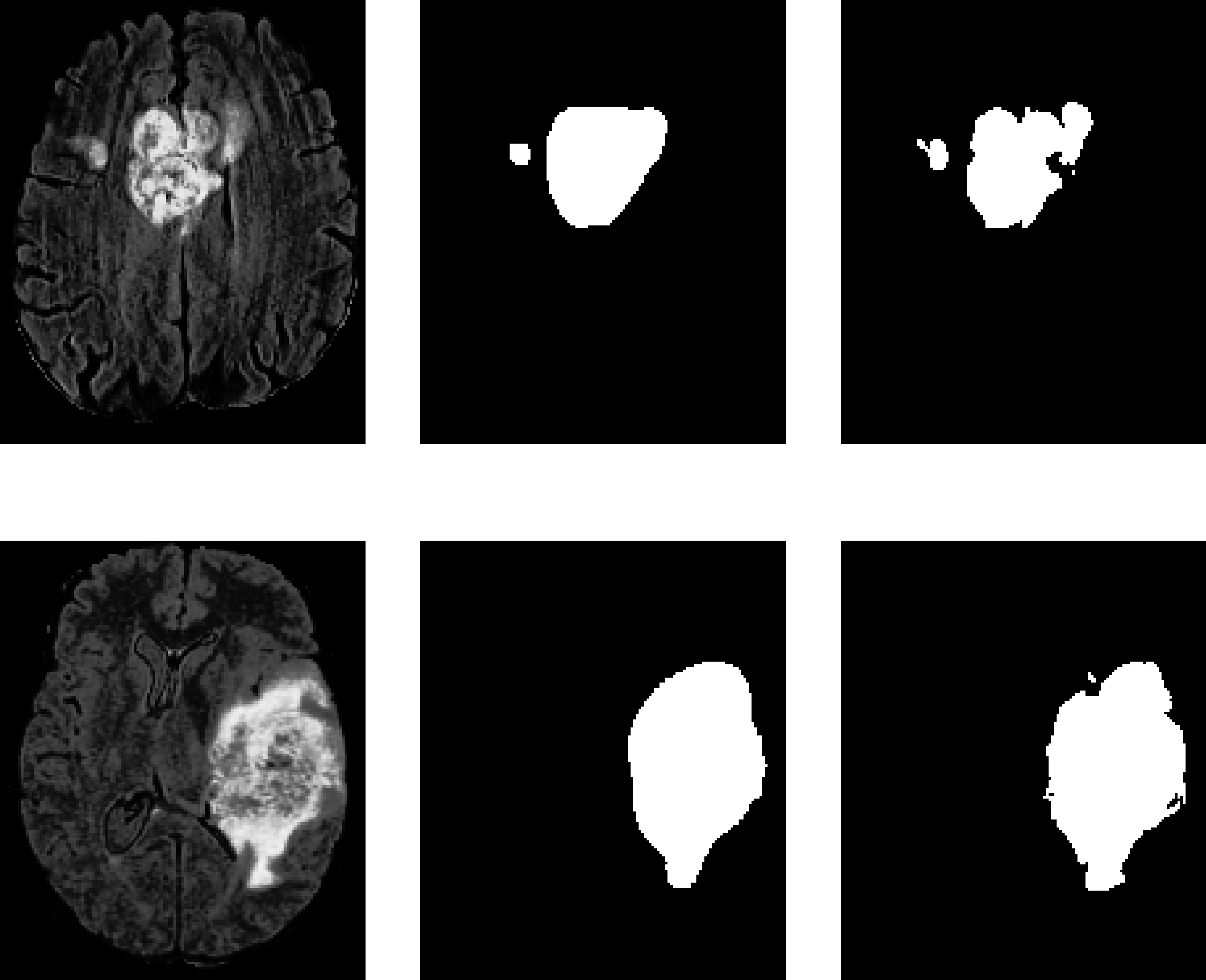}
  \label{subfig:brats-segmentation-comparison}
  }\hfill
  \subfloat[Examples from PLGNT dataset.]{%
  \includegraphics[width=0.45\linewidth]{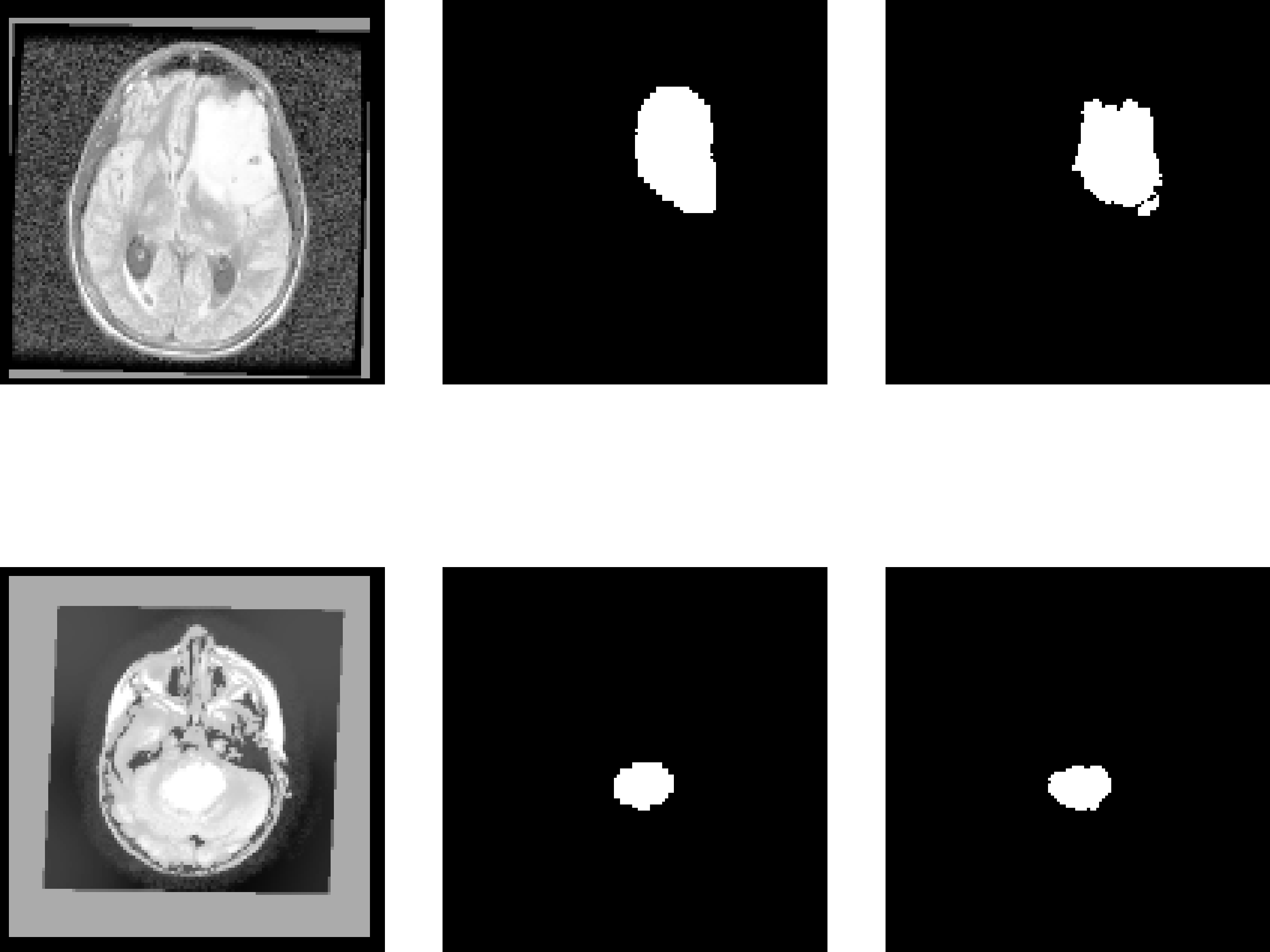}
  \label{subfig:plgnt-segmentation-comparison}
  }
  \caption{MR image FLAIR channel (left), model segmentation (middle), true segmentation (right).}
  \label{fig:segmentation_comparison}
\end{figure}

Examples of cancerous MR images and their generated non-cancerous variants are presented in Figure \ref{fig:conversion}. Examples of cancerous MR images, their generated segmentations, and their true segmentations are visualized in Figure \ref{fig:segmentation_comparison}.

We evaluated how the mean Dice changes with different threshold values when using our proposed WSCM to filter generated segmentations. We also compared the results of our segmentation models with two baseline methods. The first is the model trained solely using the seed loss. The second is a method done by the state-of-the-art method for 2D weakly supervised glioma segmentation on MR images, which takes the L1 distance between the FLAIR channel of an image and its non-cancerous variant, applies a median filter with a 5x5x5 filter, and then thresholds the filtered L1 distance map \cite{10.1007/978-3-030-11723-8_16}. The Dice for each baseline model and the proposed segmentation model at different WSCM thresholds is presented in Table \ref{tab:mean_dice_all}. The table also presents the percentage of images removed across all cohorts. The WSCM thresholds above 0.4 and 0.1 for BraTS and PLGNT respectively have not been included because the Dice coefficients stagnate at these thresholds. 

\begin{table}[h!]
  \caption{Mean Dice at different WSCM filtering thresholds and the percentage of images removed at each threshold for the baseline methods and the proposed method. All results are averaged across 5 separately trained models. T refers to the WSCM filtering threshold used.}
  \label{tab:mean_dice_all}
  \centering
  \resizebox{\columnwidth}{!}{%
  \begin{tabular}{clcccc}
    \toprule
    \multirow{2}{*}[-2pt]{\makecell{Dataset}}     & \multirow{2}{*}[-2pt]{\makecell{Method}}     & \multicolumn{3}{c}{Dice (\%)} & \multirow{2}{*}[2pt]{\makecell{Percentage of \\ Test Images \\Removed}} \\
    \cmidrule(lr){3-5}
    & & Training & Validation & Test \\
    \cmidrule(lr){1-1} \cmidrule(lr){2-2} \cmidrule(lr){3-3} \cmidrule(lr){4-4} \cmidrule(lr){5-5}  \cmidrule(lr){6-6}
    & Baseline (L1) & 55.74 &  58.05 & 56.39 & - \\
    & Baseline (Seed) & 65.01 & 66.64 & 61.15 & - \\
    & Proposed (T=0.0) & 72.48 & 73.55 & 70.29 & - \\
    BraTS & Proposed (T=0.1) & 75.60 & 76.43 & 73.26 & 5.23 \\
    & Proposed (T=0.2) & 80.41 & 80.43 & 79.67 & 18.80  \\
    & Proposed (T=0.3) & 84.15 & 83.20 & 83.91 & 28.87 \\
    & \textbf{Proposed (T=0.4)} & \textbf{86.79} & \textbf{85.78} & \textbf{86.74} & 40.58 \\
    \midrule
    \multirow{4}{*}{PLGNT} & Baseline (L1) & 62.58 & 54.90 & 57.32 & - \\
    & Baseline (Seed) & 72.43 & 54.82 & 58.48 & - \\
    & Proposed (T=0.0) & 75.93 & 60.89 & 64.06 & - \\
    & \textbf{Proposed (T=0.1)} & \textbf{80.28} & \textbf{64.67} & \textbf{68.62} & 11.41 \\
    \bottomrule
  \end{tabular}}
\end{table}

\begin{figure}[!htbp]
  \centering
  \includegraphics[width=0.8\linewidth]{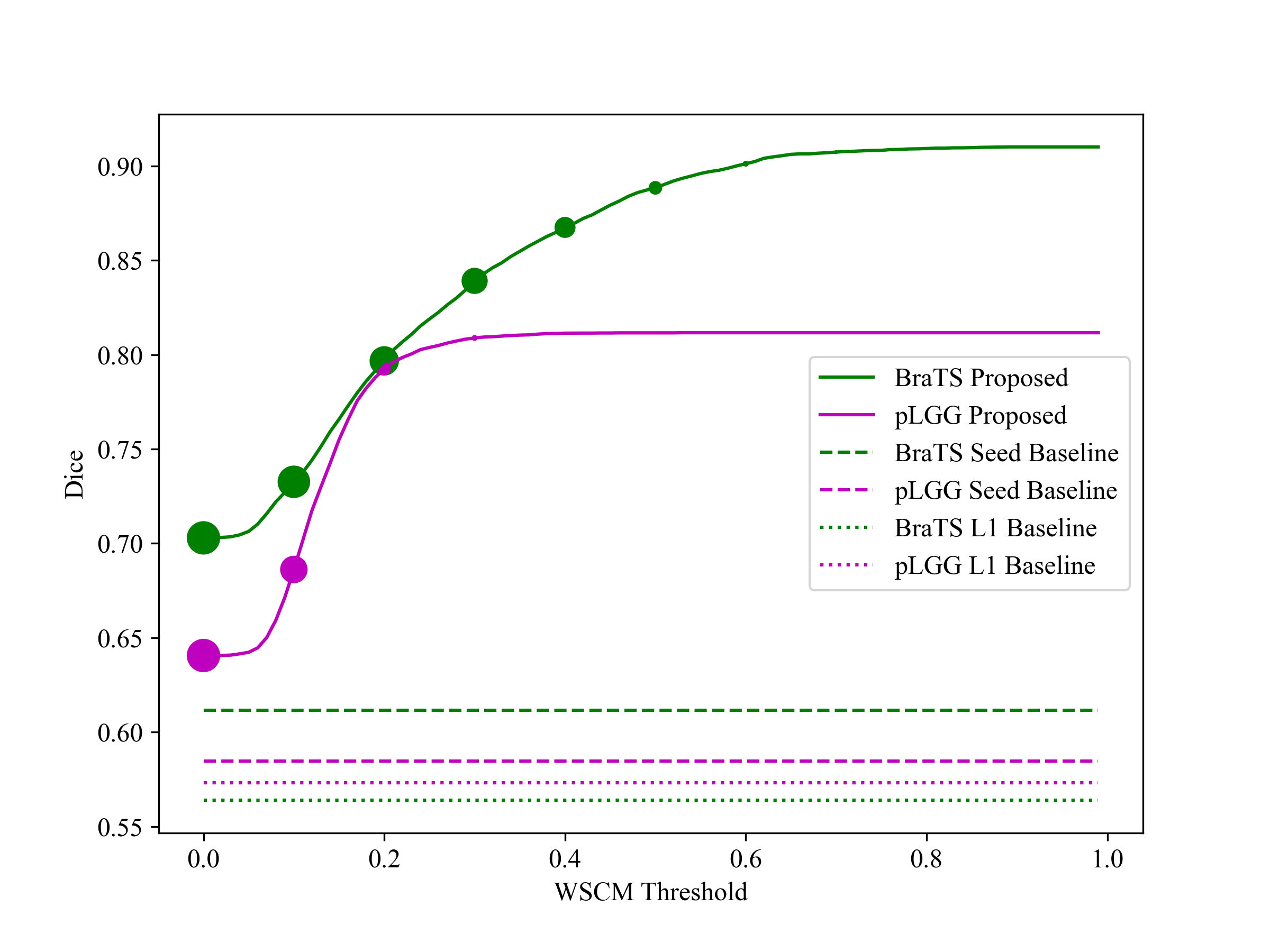}
  \caption{Increase in Dice as WSCM threshold increases for test cohort. Marker sizes indicate percentage of segmented images that are removed at each 0.1 interval of the WSCM threshold. The dashed line indicates the Dice of the seed-based baseline.}
  \label{fig:segmentation_model_system}
\end{figure}

\subsection{Radiomics-based Classification Performance}
\begin{figure}[!htbp]
  \centering
  \includegraphics[width=0.6\linewidth]{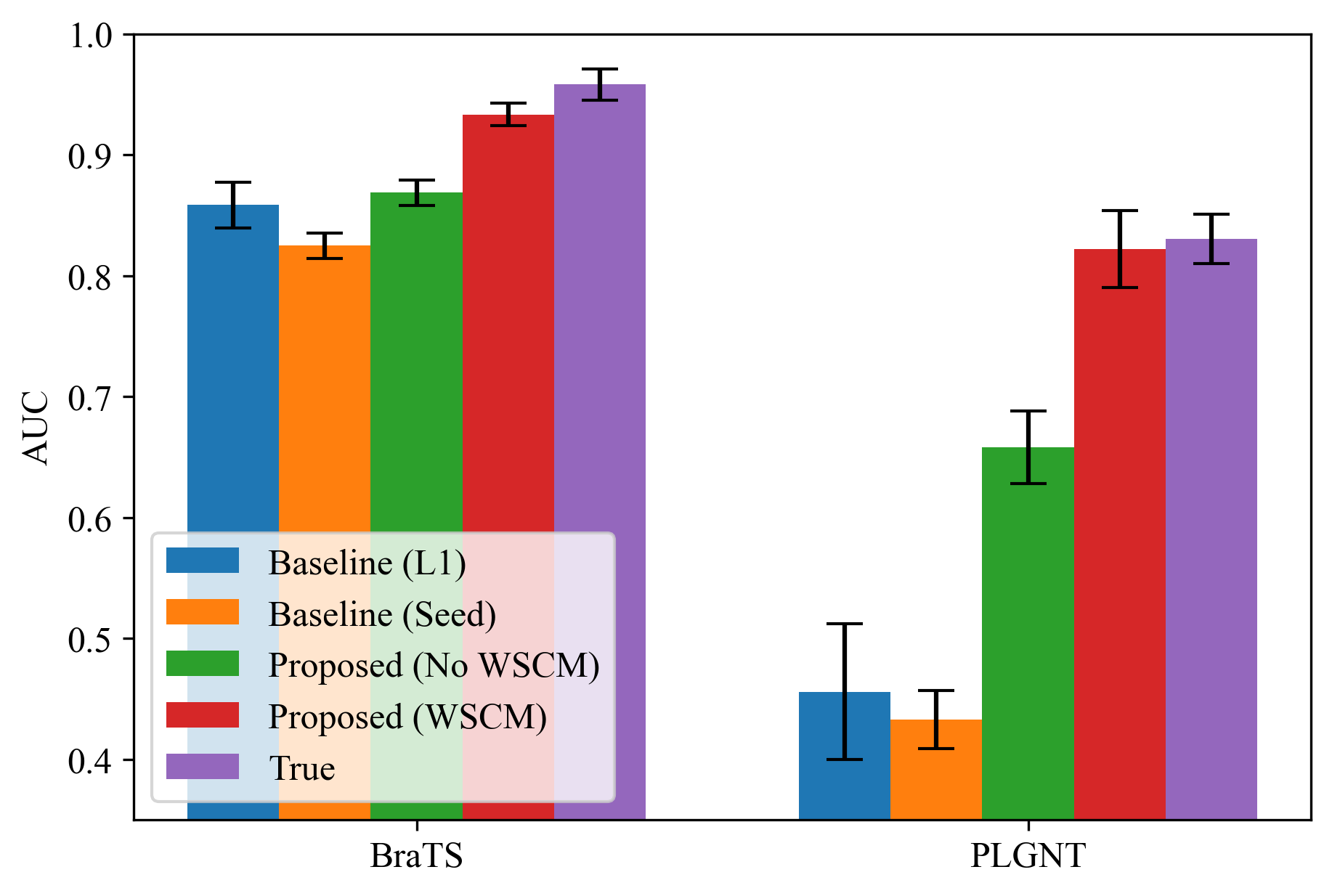}
  \caption{Bar graphs of AUCs for tumor classifiers trained using radiomic features extracted from baseline segmentations, proposed segmentations without WSCM, proposed segmentations with WSCM, and true segmentations.}
  \label{fig:auc_bar_graphs} 
\end{figure}

The random forest models were trained to predict the pathology (low-grade glioma vs high-grade glioma) for the BraTS dataset and the genetic markers (BRAF fusion vs BRAF V600E mutation) for the PLGNT dataset. The hyperparameters of the random forest models were determined using grid-search based on the best validation AUC, and the decision threshold was determined by using Youden's J statistic to acquire the optimal probability threshold for the validation cohort. For weakly supervised segmentations, 5 slices were selected from each volume using the method detailed in Section \ref{sec:method_radiomics}, and their predictions were then averaged. For the true segmentations, the 5 largest segmentations from each volume were used, and their predictions were then averaged. The test AUCs, sensitivities, and specificities of random forest models trained using Radiomic features extracted from the weakly supervised segmentations and the true segmentations for both datasets are presented in Table \ref{tab:brain_radiomics_weaklysupervised_results}. Figure \ref{fig:auc_bar_graphs} visually compares the AUCs of the classification models trained on radiomic features extracted from different segmentations. 
It can be seen that using the weakly supervised segmentations produced models with AUCs within 3\% of the models trained on the true segmentations. 

\begin{table}[htbp]
  \caption{Test mean AUC, specificity, and sensitivity for random forest models trained on radiomic features extracted from selected weakly supervised slice segmentations and true slice segmentations.}
  \label{tab:brain_radiomics_weaklysupervised_results}
  \centering
  \resizebox{\columnwidth}{!}{%
  \begin{tabular}{llccc}
    \toprule
    Dataset & Segmentations & AUC (\%) & Sensitivity (\%) & Specificity (\%) \\
    \cmidrule(lr){1-1} \cmidrule(lr){2-2} \cmidrule(lr){3-3} \cmidrule(lr){4-4} \cmidrule(lr){5-5}
    & Baseline (L1) & $85.84$ & $95.44$ & $17.88$ \\
    & Baseline (Seed) & $82.48$ & $83.85$ & $69.18$ \\
    BraTS & Proposed (No WSCM) & $86.87$ & $90.53$ & $71.42$ \\
    & Proposed (WSCM) & $93.32$ & $75.84$ & $87.16$ \\
    & True & $95.80$ & $86.72$ & $89.40$ \\
    \midrule
    & Baseline (L1) & $45.58$ & $66.33$ & $46.00$ \\
    & Baseline (Seed) & $43.26$ & $44.50$ & $45.23$ \\
    PLGNT & Proposed (No WSCM) & $65.80$ & $60.40$ & $65.10$ \\
    & Proposed (WSCM) & $82.19$ & $59.30$ & $74.42$ \\
    & True & $83.03$ & $48.50$ & $91.67$ \\
    \bottomrule
  \end{tabular}}
\end{table}

\section{Discussion and Conclusion}

Our proposed method produces segmentations with Dice coefficients greater 70\% and 64\% on the BraTS and PLGNT datasets respectively, which can then be further improved to 86\% and 68\% by filtering segmentations using the proposed WSCM. We have also demonstrated that our proposed method significantly outperforms the baseline methods, including one that is based on the state-of-the-art method for weakly supervised segmentation on 2D MR brain images \cite{10.1007/978-3-030-11723-8_16}. 

The proposed WSCM approximates the quality of the segmentations in a weakly supervised manner. It can then be used to identify a set of images that may require manual segmentation and thus relieve radiologists of the workload required to manually segment all MR images. The WSCM can also be used to identify weakly supervised segmentations that are as effective as true segmentations from each patient volume, which can then be used for Radomics-based tumor classification. This downstream clinical classification task using weakly supervised segmentations yielded performance within 3\% of the performance when using the true segmentations. 

One notable limitation of the proposed method is its reliance on the initial classifier. Poor classifier performance will negatively affect the downstream segmentation performance. In addition, this work was only applied to 2D brain MR images, and could benefit from further investigation in 3D applications and other modalities. 

We present a weakly supervised method for segmenting brain tumors in 2D MR images using a combination of localization seeds and non-cancerous variant MR images generated using GANs. Our method surpasses state-of-the-art methods for weakly supervised segmentation on 2D MR brain images. The non-cancerous variants can also be used to evaluate the segmentations in a weakly supervised manner, reducing the risk of using flawed segmentations and enabling use in downstream clinical tasks. We demonstrated the effectiveness of our weakly supervised segmentations on tumor classification tasks using two datasets, demonstrating the capability of our segmentations and how the WSCM can be used to effectively apply weakly supervised segmentations in clinical settings. 

\section*{Acknowledgements}
\label{sec:acknowledgements}
This work was supported in part by Huawei Technologies Canada Co., Ltd. and Chair in Medical Imaging and Artificial Intelligence, a joint Hospital-University Chair between the University of Toronto, The Hospital for Sick Children, and the SickKids Foundation.

\end{document}